%% file: main.tex
\documentclass[preprint,10pt,authoryear]{elsarticle}
\usepackage[british]{babel}
\usepackage[a4paper,margin=2cm]{geometry}
\usepackage{lineno}
\usepackage{hyperref}
\hypersetup{colorlinks=true,linkcolor=blue,citecolor=blue,urlcolor=blue}
\usepackage{amssymb,amsmath}
\usepackage{svg}
\usepackage{graphicx}
\usepackage{tabularx,booktabs,array,ragged2e,threeparttable}
\usepackage{caption,subcaption}
\usepackage{float}
\usepackage{xargs}
\usepackage[disable,prependcaption]{todonotes}

\newcommandx{\linus}[2][1=]{}
\newcommandx{\ben}[2][1=]{}
\newcommandx{\sanja}[2][1=]{}
\newcommandx{\stephen}[2][1=]{}
\newcolumntype{L}[1]{>{\RaggedRight\arraybackslash}p{#1}}
\journal{Health \& Place}
\begin{document}
\begin{frontmatter}

\title{Neighbourhood Walkability and Cardiometabolic Morbidity in Greater London: A Fine-Grained Spatial Analysis}

\author[kcl]{Benjamin T. Pike}
\author[cuhk]{Stephen Law}
\author[nokia]{Sanja Šćepanović}
\author[kcl]{Linus W. Dietz}
\affiliation[kcl]{organization={King's College London}, city={London}, country={United Kingdom}}
\affiliation[cuhk]{organization={The Chinese University of Hong Kong}, city={Hong Kong}, country={Hong Kong}}
\affiliation[nokia]{organization={Nokia Bell Labs}, city={Cambridge}, country={United Kingdom}}

\begin{abstract}
\input{sections/abstract.tex}

\end{abstract}

\end{frontmatter}

\section{Introduction}
\label{sec:introduction}
\input{sections/introduction.tex}

\section{Methods}
\label{sec:methods}
\input{sections/methods.tex}

\section{Results}
\label{sec:results}
\input{sections/results.tex}

\section{Discussion}
\label{sec:discussion}
\input{sections/discussion.tex}

\section{Conclusion}
\input{sections/conclusion.tex}

\section*{Acknowledgements}
We thank Ed Parham and Space Syntax Limited for sharing the Space Syntax 15-minute walkability index.

\bibliographystyle{elsarticle-harv}
\bibliography{literature.bib}
\end{document}

%% file: sections/abstract.tex
Cardiometabolic diseases are leading causes of ill health, and neighbourhoods that support everyday walking may help reduce them. Studies often report lower disease prevalence in more walkable neighbourhoods. However, because walkability operates across multiple spatial scales, city-wide associations may conflate local neighbourhood form with broader metropolitan structure and co-varying environmental exposures. Here we show that while much of the walkability--morbidity association in Greater London operates through the city's macro-scale spatial structure, local alignment persists beyond it. Using an open five-component walkability index for 4,994 London census units and recorded prevalence of eight cardiometabolic conditions from primary-care registers, we found that higher walkability was inversely associated with all eight outcomes after adjusting for demographic composition and deprivation; one standard deviation higher walkability corresponded to 1.43 percentage points lower hypertension prevalence. Walkability followed a centre--periphery gradient, with a radial model explaining 74.2\% of its variation. Coefficients were reduced by 2--43\% within boroughs and 11--75\% after centre--periphery adjustment. After spatial-error adjustment, no clear association remained for obesity; coefficients for the other seven were reduced by 83--99\% relative to comparable log-scale models. Nevertheless, observed associations were more extreme than those from surrogate walkability surfaces preserving this spatial structure, for all eight outcomes in the primary analyses and for five after adjustment for nitrogen oxides. These findings caution against interpreting city-wide walkability coefficients as neighbourhood effects independent of wider metropolitan structure, and highlight the need for studies of neighbourhood form and health to disentangle global and local spatial structure explicitly.

%% file: sections/introduction.tex
Cardiometabolic diseases are among the leading contributors to global morbidity and mortality \citep{GBD2024}. Their development is strongly influenced by modifiable behaviours, particularly physical inactivity and sedentary behaviour \citep{Bell2022}. Walking is among the most accessible forms of physical activity, but whether it features in everyday life depends partly on the environments through which people move. Neighbourhoods that place homes, services, destinations, and public transport within practical walking distance, connected through direct and permeable street networks, reduce barriers to routine active travel \citep{Cervero1997,Ewing2010}%
. Neighbourhood design has consequently become an important consideration in cardiometabolic health \citep{Koohsari2020,Adhikari2021,Frank2022}%
.

The extent to which a neighbourhood supports walking is commonly described as its \emph{walkability}. Rather than representing a single physical characteristic, walkability captures several dimensions of urban form. The original ``3 Ds'' framework emphasised density, diversity of land use, and street-network design \citep{Cervero1997,Frank2009}%
, while subsequent formulations have incorporated destination accessibility and public transport, producing the widely used ``5 Ds'' framework \citep{Ewing2010,Venerandi2024}. These dimensions influence walking through related but distinct mechanisms: density shortens travel distances; mixed land use reduces spatial segregation; interconnected streets provide direct routes; and accessible public transport extends practical walking range and reduces car dependency \citep{Venerandi2024}. Contemporary indices therefore tend to combine several characteristics rather than relying on any single measure \citep{Stockton2016,Lam2022,Venerandi2024}.%

There remains, however, no agreed method for translating these concepts into a walkability measure. Early indices derived using geographic information systems combined dwelling density, land-use mix, and intersection density, while newer measures incorporate pedestrian infrastructure, greenery, traffic conditions, and perceived safety \citep{Frank2005,Frank2009,TalaveraGarcia2015,Alves2020}%
. National and regional measures in the United States, Canada, and the Netherlands demonstrate that objective indices can be produced consistently over large areas \citep{Thomas2017,Herrmann2019,Lam2022}%
. Yet existing indices vary substantially in constituent variables, weighting schemes, and definitions of neighbourhood exposure \citep{Glazier2012,Brown2023}%
. Greater complexity also increases data requirements and can reduce transferability \citep{Maghelal2011,Shashank2019}%
, while commercial measures such as Walk Score rely on proprietary methods that limit reproducibility \citep{Carr2010,Herrmann2017}%
. Many indices also aggregate environmental characteristics within administrative units or Euclidean buffers despite pedestrian movement being constrained by street networks and physical barriers \citep{Ellis2015,Venerandi2024}, which makes them sensitive to the spatial unit chosen \citep{Maghelal2011,Shashank2019}. An established index for London illustrates these constraints: it combines residential density, street connectivity, and land-use mix within census areas \citep{Stockton2016}, and therefore omits destination and public-transport accessibility.

Epidemiological evidence nevertheless suggests that higher walkability is associated with lower cardiovascular morbidity and mortality \citep{Lang2022,Koohsari2020,Jia2018}%
, hypertension \citep{Sarkar2018,Jones2021,Chen2024}%
, adiposity and obesity \citep{Frank2022,Rundle2023}%
, and diabetes \citep{Frank2022,Hua2024}%
. Findings are not uniform, however: a population-based cohort study reported weaker or null associations for some cardiovascular outcomes \citep{Meijer2025}, and relationships vary across populations and geographical settings \citep{Chandrabose2019,Kartschmit2020,Cerin2024}%
. This variation suggests that the health implications of walkability cannot be understood solely from neighbourhood form.

Walkability is also embedded within a wider social and environmental context. Dense, well-connected neighbourhoods may simultaneously experience socioeconomic disadvantage or greater traffic-related air pollution, while lower-density areas may offer greater vegetation and open space. Socioeconomic conditions can modify or confound walkability--health associations \citep{Lang2022,Koohsari2020,Adhikari2021}%
; traffic-related pollution may offset some benefits associated with active travel \citep{Howell2019,Zhang2025,Zhu2023}%
; and greenness may provide complementary benefits \citep{Scheer2024,Dzhambov2025}. Co-occurring neighbourhood characteristics may therefore be only attributed to walkability if the wider urban environment is not considered. Residential self-selection and incomplete measurement of activity spaces further constrain causal interpretation \citep{Shashank2019,Shields2021}%
.

These challenges are compounded by the spatial nature of both exposure and outcome. Walkability is strongly clustered because urban form develops continuously across neighbouring areas, and London's centre--periphery gradient in walkability has been documented previously \citep{Stockton2016}. Similar gradients are seen in other urban characteristics, such as urban vibrancy and park quality \citep{Scepanovic2021,Dietz2025Examining}. Cardiometabolic morbidity is also geographically patterned through shared demographic, socioeconomic, and environmental determinants. When exposure and outcome both follow broad geographical patterns, a city-wide association may partly reflect where neighbourhoods are located rather than how walkable they are, because unmeasured influences on health that vary smoothly across the city are confounded with location. Conventional regression conflates these geographical contrasts. It can overstate precision when spatial dependence is not represented, and the estimated coefficient itself can change substantially once spatial structure is modelled.

Previous work has demonstrated sensitivity to neighbourhood definitions and geographical variation in relationships between neighbourhood characteristics and health \citep{Coffee2013,Xu2023}. Less attention has been given to how much of an observed city-wide association is shared with the broader spatial organisation of the metropolitan area, and how much arises from the local alignment between walkability and morbidity. A further open question is whether an association of similar size would arise from any exposure with the same spatial structure. Spatially constrained null models address this by comparing an observed association with those generated by surrogate surfaces that preserve the exposure's spatial structure.

The growing availability of geographic data now makes fine-grained measurement increasingly feasible\citep{Boeing2025}. OpenStreetMap provides detailed coverage of streets, footpaths, and amenities, %
while national datasets provide residential, public-transport, environmental, and health information. Composite resources such as MedSat further consolidate neighbourhood-level demographic and health measures \citep{Scepanovic2023}. Greater London provides a useful setting because its 4,994 Lower Layer Super Output Areas (LSOAs) capture substantial variation in urban form, environmental conditions, and socioeconomic composition within a single metropolitan region.

In this study, we examine not only whether walkability is associated with cardiometabolic morbidity in Greater London, but also how the estimated association depends on the wider spatial structure of the city. To measure walkability, we develop an open and reproducible index with five components: dwelling density, land-use mix, street connectivity, amenity accessibility, and public-transport accessibility. Unlike indices based on administrative units, we calculate each component with multiple 15-minute network-derived walksheds. We examine the index's spatial structure and component composition, validate it against observed travel behaviour and two existing London walkability measures (the index of \citet{Stockton2016} and the Space Syntax 15-minute walkability index), and estimate associations with recorded prevalence of eight cardiometabolic conditions. Air pollution and urban greenness are examined separately as contextual environmental exposures. To assess the role of London's wider geography, we compare city-wide estimates with those obtained within boroughs and from models accounting for the centre--periphery gradient and spatial dependence. We also examine whether similar associations arise from spatially constrained surrogate walkability surfaces that preserve this spatial structure. Further analyses assess alternative index definitions, neighbourhood scales, non-linearity, influential observations, and predictive performance under spatial-block cross-validation.

%% file: sections/methods.tex
\subsection{Study area and neighbourhood walksheds}
We conducted a cross-sectional ecological study across the 4,994 LSOAs from the 2021 Census that fell within the administrative boundary of Greater London. LSOAs had a mean area of approximately 0.32~km$^2$ and mean population of approximately 1,800. All spatial calculations used the British National Grid projection (EPSG:27700).

A pedestrian network covering Greater London was derived from OpenStreetMap (OSM) using the Python package OSMnx \citep{OpenStreetMap2025,Boeing2025} in October 2026. %
 Paths closed to pedestrians were excluded, isolated fragments with fewer than 500 nodes were removed, and the network was simplified, with every path treated as walkable in both directions. The resulting graph contained 571,537 nodes and 745,332 edges.

Ten \emph{seed} nodes were assigned to each LSOA. Where fewer than ten candidate nodes were available, the search area was expanded until at least ten were obtained. Candidate nodes were then grouped into ten spatial clusters using $k$-means clustering of their coordinates, and the node nearest the centre of each cluster was selected. This spread the seeds across each LSOA instead of representing the neighbourhood by a single centroid.

From each seed, we identified every part of the network reachable within a 15-minute walk using Dijkstra's shortest-path algorithm. Walking times were calculated from path length assuming a walking speed of 1.33~m/s (approximately 4.8~km/h), consistent with estimates of habitual adult outdoor walking speed \citep{Murtagh2020}. A delay of 1.6~s was added at every junction where three or more paths meet, to approximate crossing delays \citep{Schultz2019}.

The reachable paths were widened by 25~m on each side to capture street-fronting buildings and destinations, and merged into a single area for each seed, its \emph{walkshed}. The ten walksheds of each LSOA were kept separate, with component measures calculated for each and then averaged. This preserved variation in accessibility from different locations within each LSOA and avoided constructing a single catchment representing the combined reach of multiple origins.

\subsection{Composite walkability index}

Walkability was quantified using five components, one for each dimension of the ``5 Ds'' framework \citep{Ewing2010}, using contemporary terminology as \emph{amenity} accessibility, \emph{dwelling} density, \emph{land-use} mix, \emph{street} connectivity, and public-\emph{transport} accessibility. This extends the density, connectivity, and land-use components of the existing London index \citep{Stockton2016} with destination and public-transport accessibility, while relying only on open data. Each component score was standardised to a z-score across the study area and the composite score for LSOA $i$ was calculated as their unweighted mean,
\begin{equation}
W_i=\frac{1}{5}\left(
z_i^{(\mathrm{amenity})}
+z_i^{(\mathrm{dwelling})}
+z_i^{(\mathrm{landuse})}
+z_i^{(\mathrm{street})}
+z_i^{(\mathrm{transport})}
\right).
\end{equation}

Because existing indices differ in their weighting schemes \citep{Glazier2012,Brown2023}, equal weighting was used to preserve interpretability and avoid imposing an unvalidated outcome-specific weighting scheme. The resulting composite had an empirical standard deviation (SD) of 0.773 score units, which was used to express associations per 1-SD increase in walkability.

\subsubsection{Amenity accessibility}

Amenity accessibility was defined using seven destination categories relevant to everyday pedestrian activity: essentials, eating out, education, healthcare, culture, leisure, and services. A separate \emph{vice} category captured destinations such as fast-food outlets, pubs, bookmakers, and tobacco shops. These were retained for analysis but excluded from the walkability score.

Points of interest were taken from OSM, whose tagging model usies free-form key--value tags rather than a fixed taxonomy. Tags occurring at least 100 times globally were retrieved from \href{https://taginfo.openstreetmap.org/tags}{OSM Taginfo}, yielding 1,661 candidates, and classified using a large language model (GPT-4o, temperature 0.3) into the eight destination categories or as non-amenity features. A random 10\% sample was manually reviewed for face validity.   %
The final mapping retained 620 tags: services (152), leisure (138), culture (111), essentials (90), healthcare (63), education (28), eating out (14), and vice (24).

Features matching the seven supportive categories were counted within each walkshed. Counts were averaged across the ten seeds, transformed using $\log(1+x)$ to reduce right skew, and standardised. Amenity accessibility was calculated as the unweighted mean of the seven category z-scores,
\begin{equation}
A_i=\frac{1}{7}\sum_{c\in C} z_i^{(c)},
\end{equation}
where $C$ denotes the seven supportive amenity categories.

\subsubsection{Dwelling density}

Dwelling density, a core component of established walkability indices \citep{Frank2005,Frank2009,Stockton2016}, was calculated using the domestic Energy Performance Certificate register as a proxy. Its records represent the energy performance for individual dwellings, including separate flats within multi-unit buildings, and were preferred to OSM building footprints because residential coverage and dwelling counts are more consistent. Each record was located using the Ordnance Survey coordinates of its Unique Property Reference Number. Within each walkshed, dwelling density was calculated as the number of accessible dwellings divided by walkshed area,
\begin{equation}
D_i^{(s)}=\frac{n_i^{(s)}}{a_i^{(s)}}.
\end{equation}

Density estimates were averaged across the ten walksheds to produce the dwelling-density value for each LSOA.

\subsubsection{Land-use mix}

Land-use mix was quantified using the \emph{OSM Landuse Landcover} map, a grid of 10-m cells \citep{Schultz2024}. This map combines land use mapped in OSM with a random-forest classifier trained on Sentinel-2 satellite imagery for otherwise unmapped areas, providing continuous coverage across the study area. The November 2024 tiles covering Greater London were clipped to the study boundary. Its 14 land-cover classes (CORINE level 2) were merged into five functional groups: \emph{residential}; \emph{industrial}, including commercial and transport land; \emph{greenspace}; \emph{bluespace}; and \emph{disturbance}, including construction and other non-vegetated land. This reduced sensitivity to distinctions of limited relevance to walking behaviour \citep{Manaugh2013}.

For each walkshed $s$ in LSOA $i$, the proportion of cells in each group, $p_{gi}^{(s)}$, was calculated. Land-use mix was quantified as the evenness of these proportions, using normalised Shannon entropy,
\begin{equation}
L_i^{(s)}
=
-\frac{1}{\ln 5}
\sum_{g=1}^{5}
p_{gi}^{(s)}\ln p_{gi}^{(s)}.
\end{equation}

The resulting score ranges from 0, when all classified cells belong to a single group, to 1, when cells are distributed equally across all five groups. Entropy values were averaged across the ten walksheds to produce the land-use-mix value for each LSOA.

\subsubsection{Street connectivity}

Street connectivity was characterised using seven measures of the street network. Higher intersection density \citep{Frank2005,Stockton2016,TalaveraGarcia2015}, street-length density \citep{Tresidder2005}, and mean intersection degree (the mean number of streets meeting at an intersection; \citealp{Berrigan2010}) indicated denser, more interconnected networks. Conversely, four measures indicated lower connectivity when higher: circuity (greater route indirectness; \citealp{Boeing2019}), dead-end density (barriers to through-movement; \citealp{Molaei2021}), mean edge betweenness centrality (concentration of shortest paths; \citealp{Kirkley2018,Zaleckis2022}), and mean block length (coarser network granularity; \citealp{Brown2023,Hirsch2013}).

Street structure was measured using OSM street networks clipped to each walkshed. Restricting the analysis to streets reduced sensitivity to inconsistent sidewalk mapping in OSM, which can substantially alter network structure and the resulting connectivity measures. Networks were simplified before the measures were calculated, and those with fewer than 60 nodes were excluded. Intersection density, street-length density, mean intersection degree, circuity, and dead-end density were calculated using OSMnx; mean edge betweenness centrality using the igraph library; and mean block length using a custom function measuring network distance between consecutive intersections. LSOA-level values were obtained by averaging the valid estimates across walksheds. Measures were standardised and oriented so that higher values consistently represented greater connectivity, then combined as an unweighted mean,
\begin{equation}
S_i=\frac{1}{7}\sum_{m=1}^{7} z_i^{(m)},
\end{equation}
where $z_i^{(m)}$ denotes the standardised, directionally oriented value of measure $m$ for LSOA $i$.

\subsubsection{Public-transport accessibility}

Public-transport accessibility incorporated both the availability of reachable stops and the level of service they provide. Bus timetables were obtained from the \href{https://data.bus-data.dft.gov.uk/downloads/}{Bus Open Data Service}, rail timetables from \href{https://www.networkrail.co.uk/who-we-are/transparency-and-ethics/transparency/open-data-feeds/}{Network Rail}, and London Underground, Docklands Light Railway, river bus, and tram timetables from \href{https://tfl.gov.uk/info-for/open-data-users/api-documentation}{Transport for London}. Timetables not supplied in the General Transit Feed Specification, the standard open timetable format, were converted to it using the UK2GTFS tool \citep{Morgan2025}.

Mean daily service frequency, $f_j$, was calculated for each stop across the seven-day week. To represent broad differences in carrying capacity and travel speed, each mode was assigned heuristic capacity ($c_m$) and speed ($v_m$) weights. These were intended to represent relative differences between modes rather than precise vehicle capacities or operating speeds. The raw utility of stop $j$ was defined as
\begin{equation}
u_j = f_j c_{m(j)} v_{m(j)},
\end{equation}
where $m(j)$ denotes the transport mode serving stop $j$.

Because transport utility was strongly right-skewed and increases in service provision were not assumed to confer proportionally increasing accessibility, a diminishing-returns transformation was applied. This reduced the influence of exceptionally high-service locations while preserving their ordering,
\begin{equation}
w_j = \log\left(\sqrt{u_j}+1\right).
\end{equation}

Transformed stop utilities were summed within each walkshed,
\begin{equation}
T_i^{(s)} = \sum_{j \in P_i^{(s)}} w_j,
\end{equation}
where $P_i^{(s)}$ denotes stops within walkshed $s$ of LSOA $i$. Scores were averaged across the ten walksheds to produce the LSOA-level value.

\subsection{Recorded cardiometabolic prevalence}

Chronic disease prevalence is not published at LSOA level. The finest-grained available data are reported for general practices (GPs), whose patients are drawn from many LSOAs. We therefore adapted the spatial-allocation approach introduced in MedSat \citep{Scepanovic2023}, redistributing Quality and Outcomes Framework (QOF) disease registers according to the residential distribution of registered patients. Eight registers were analysed: hypertension, diabetes mellitus, obesity, coronary heart disease (CHD), atrial fibrillation (AF), heart failure (HF), peripheral arterial disease (PAD), and stroke/transient ischaemic attack (TIA).

\subsubsection{Data sources and preparation}

Annual QOF data for the financial years 2020/21--2025/26 provided condition-specific practice prevalence. For practice $p$, condition $c$, and year $t$, the reported prevalence was denoted
\begin{equation}
r_{pc}^{(t)}
=
\frac{\text{patients on register }c}
{\text{registered practice population}}.
\end{equation}

The April editions for 2021--2026 of NHS England's \emph{Patients Registered at a GP Practice} dataset provided the number of patients registered at practice $p$ who resided in LSOA $i$, denoted $n_{pi}^{(t)}$. Registrations reported using 2011 LSOAs were mapped to 2021 LSOAs using the Office for National Statistics exact-fit lookup. Counts were distributed across split LSOAs in proportion to their 2021 Census populations and summed where LSOAs had merged, assuming consistent practice registration patterns within each original LSOA.

\subsubsection{Spatial allocation of recorded prevalence}

Assuming that patients on each disease register follow the same residential distribution as the practice population as a whole, annual recorded prevalence for condition $c$ in LSOA $i$ was estimated as
\begin{equation}
\pi_{ic}^{(t)}
=
\frac{\sum_{p\in P_c^{(t)}} n_{pi}^{(t)}r_{pc}^{(t)}}
{\sum_{p\in P_c^{(t)}} n_{pi}^{(t)}},
\end{equation}
where $P_c^{(t)}$ denotes practices with a usable QOF prevalence estimate for condition $c$ in year $t$. Thus, $\pi_{ic}^{(t)}$ is the patient-weighted mean of QOF prevalence among practices serving residents of LSOA $i$. Practices without usable estimates were excluded from both numerator and denominator.

Allocation was performed separately for each financial year and annual estimates were averaged to reduce year-to-year volatility,
\begin{equation}
\bar{\pi}_{ic}
=
\frac{1}{6}
\sum_{t=2021}^{2026}
\pi_{ic}^{(t)}.
\end{equation}

The resulting $\bar{\pi}_{ic}$ values were used as LSOA-level recorded prevalence estimates. Before modelling, the upper 0.1\% of LSOA values for each condition were set to missing to limit the influence of extreme estimates.

\subsection{Covariates and contextual environmental variables}

Demographic composition and deprivation were treated as the primary adjustment covariates. Air pollution and greenness were examined separately as contextual environmental exposures because both are closely coupled to the neighbourhood characteristics captured by the walkability index and may represent co-occurring characteristics or pathways linking urban form to health, rather than purely upstream confounders.

Demographic composition was obtained from MedSat \citep{Scepanovic2023} and included the proportion male; the proportions aged 0--14, 15--44, 45--64, and 65+ years; and the proportions of White, Asian, Black, and Mixed ethnicity. Deprivation was measured using the Index of Multiple Deprivation 2025 \citep{IMD2025}. Traffic-related air pollution was represented by annual mean nitrogen oxides (NO\textsubscript{x}) from the 2023 1-km Pollution Climate Mapping grids of the Department for Environment, Food and Rural Affairs, aggregated to LSOAs using area-weighted means. Urban greenness was represented by the mean 2023 Normalised Difference Vegetation Index (NDVI) from the Small Area UK Vegetation Indices dataset, derived from the per-pixel median of cloud-masked Sentinel-2 satellite images taken between April and October. Continuous covariates were standardised.

\subsection{Statistical analysis}

Index structure was examined using pairwise correlations, variance inflation factors, and principal component analysis. Sensitivity to construction was assessed using five leave-one-component-out indices and an alternative first principal component index, with rank agreement quantified using Spearman correlation.

External correspondence of our index was assessed against two existing London walkability measures: the index of \citet{Stockton2016}, which combines residential density, street connectivity, and land-use mix, and the 15-minute walkability index of the Space Syntax UK Integrated Urban Model, a street-segment-level measure derived from network configuration and accessibility.

Geographical patterning was described relative to the geographical centre (centroid) of the study area. Composite walkability was modelled as a continuous piecewise linear function of distance from this centre (radial distance), with breakpoints estimated at approximately 8 and 18~km. Directional variation was examined across eight $45^\circ$ compass sectors. LSOAs at least one residual standard deviation below their radial expectation were classified as local shortfalls; the component furthest below its corresponding radial expectation was designated the dominant deficit if it was also at least one residual standard deviation below expectation. Otherwise, the shortfall was classified as mixed.

Unadjusted relationships between walkability and cardiometabolic outcomes were summarised using Pearson correlations. Primary associations were estimated using ordinary least-squares regression adjusted for demographic composition and socioeconomic deprivation; ages 15--44 years and White ethnicity were reference categories. Complete-case analysis yielded 4,983 LSOAs. Estimates were expressed as percentage-point differences in recorded prevalence per empirical SD of walkability, with Benjamini--Hochberg correction for multiple testing across the eight outcomes. Sensitivity models additionally adjusted for NO\textsubscript{x}, NDVI, or both. Non-linearity was examined using natural cubic splines with four degrees of freedom and walkability quintiles. Effect modification by deprivation, NO\textsubscript{x}, and NDVI was assessed using interaction models, with the same correction applied across the 24 tests. Sensitivity to unequal error variance was assessed using heteroskedasticity-robust (HC3) standard errors, and sensitivity to influential observations using Cook's distance.

Spatial dependence was assessed by treating LSOAs that share a boundary or corner as neighbours (queen contiguity). Global Moran's $I$, a measure of spatial autocorrelation, was calculated for walkability, each outcome, and regression residuals. Spatial-error models, which allow regression errors to be correlated between neighbouring LSOAs, were fitted by maximum likelihood to log-transformed prevalence under the primary and environmental covariate specifications, while London borough fixed effects restricted comparisons to LSOAs within the same borough. Centre--periphery structure was represented by piecewise radial gradients fitted separately for each compass sector and added directly to the primary models. Finally, a spatially constrained null analysis tested whether the observed associations could be reproduced by 999 surrogate walkability surfaces: simulated maps that preserved the fitted centre--periphery pattern and approximately matched the residual distribution and global Moran's $I$. Tests were Benjamini--Hochberg corrected across the eight outcomes and repeated under the NO\textsubscript{x}, NDVI, and combined environmental specifications.

Additional sensitivity analyses examined 10-, 15-, and 20-minute walksheds and uncertainty in QOF spatial allocation. QOF allocation sensitivity used 100 Monte Carlo simulations in which within-practice allocation shares were varied by up to $\pm20\%$ and renormalised, while practice case totals and registered-patient denominators were held fixed. Finally, five-fold spatial-block cross-validation, in which models are tested on geographically separate blocks of LSOAs withheld from fitting, compared demographic- and deprivation-adjusted models with and without walkability, with changes in held-out $R^2$ used to assess the incremental predictive contribution of walkability in geographically held-out areas.

%% file: sections/results.tex
\subsection{Validation and characteristics of the walkability index}

Direct measures of walking behaviour are not routinely available at the LSOA-scale. External correspondence was therefore assessed using 2021 Census commuting mode shares. Higher walkability was associated with a lower share commuting by car (Spearman $\rho=-0.83$) and higher shares commuting on foot ($\rho=0.59$) or on foot and by bicycle combined ($\rho=0.74$; all $p<0.001$). These measures capture commuting rather than total walking behaviour and exclude walking undertaken as part of a public-transport journey. The index was also positively correlated with the Output Area-level London walkability index developed by \citet{Stockton2016} ($\rho=0.71$, $p<0.001$) and with the Space Syntax 15-minute walkability index ($\rho=0.85$, $p<0.001$). Table~\ref{tab:index-correlations} reports rank and linear correlations of both external measures with the composite index and each component. The Stockton index related moderately and fairly evenly to amenity accessibility, dwelling density, street connectivity, and public-transport accessibility ($\rho=0.54$--$0.70$). The Space Syntax index related strongly to the same four components ($\rho=0.72$--$0.85$) but only weakly to land-use mix ($\rho=0.11$), consistent with its derivation from street-network configuration and accessibility rather than land-use composition. Pearson correlations for the Space Syntax index were lower ($r=0.38$--$0.59$) owing to its strongly right-skewed distribution.

\input{tables/index_correlations}

Amenity accessibility, dwelling density, street connectivity, and public-transport accessibility were moderately to strongly correlated ($r=0.71$--$0.84$), together forming a broader dimension of urban intensity, whereas land-use mix was weakly related to the remaining components ($r=-0.07$ to $0.18$). Variance inflation factors ranged from 1.16 to 4.44. The first principal component explained 66.7\% of variance and largely captured this urban-intensity dimension; the second explained 20.8\% and was dominated by land-use mix.

Urban intensity was strongly associated with higher NO\textsubscript{x} ($r=0.73$--$0.80$ across its constituent components) and moderately inversely associated with NDVI ($r=-0.49$ to $-0.57$). Land-use mix showed a weaker association with NO\textsubscript{x} ($r=0.31$) and little association with NDVI ($r=-0.03$). Associations with deprivation were positive but substantially smaller ($r=0.13$--$0.25$ across components).

\subsection{Spatial patterning of walkability}

Walkability exhibited a pronounced centre--periphery gradient across Greater London (Figure~\ref{fig:spatial-anatomy}A). The decline was steepest within 8~km (0.290 SD km$^{-1}$), slower between 8 and 18~km (0.133 SD km$^{-1}$), and shallow beyond 18~km (0.028 SD km$^{-1}$). This piecewise model explained 74.2\% of variation in the composite score (Figure~\ref{fig:spatial-anatomy}C). Allowing the radial pattern to vary across the eight compass sectors increased explained variation to 80.8\%. Within 8~km, walkability was highest to the west and east and lowest to the south and south-east (Figure~\ref{fig:spatial-anatomy}D). Radial gradients were strongest for amenity accessibility and dwelling density and weakest for land-use mix, while variation between sectors was greatest for land-use mix and public-transport accessibility. Dwelling density showed a comparatively consistent radial decline.

Comparison with the piecewise radial expectation identified 811 LSOAs (16.2\%) with composite walkability at least one residual standard deviation below expectation (Figure~\ref{fig:spatial-anatomy}B). Land-use mix was the dominant deficit in 24.3\% of shortfalls, followed by amenity accessibility (22.8\%), street connectivity (20.8\%), public-transport accessibility (15.0\%), and dwelling density (14.9\%). The remaining 2.1\% had no single component at least one residual standard deviation below expectation.

\input{figures/spatial_anatomy}

\subsection{Walkability and cardiometabolic morbidity}

Higher walkability was associated with lower recorded prevalence of all eight cardiometabolic outcomes. Unadjusted correlations were strongest for hypertension ($r=-0.75$), CHD ($r=-0.70$), and stroke/TIA ($r=-0.62$), followed by AF ($r=-0.58$), HF ($r=-0.55$), diabetes ($r=-0.45$), obesity ($r=-0.38$), and PAD ($r=-0.23$).

All eight associations remained inverse after adjustment for demographic composition and socioeconomic deprivation. Across 4,983 complete-case LSOAs, a 1-SD increase in walkability was associated with 1.43 percentage points lower hypertension prevalence (95\% confidence interval, CI, $-1.48$ to $-1.37$), 0.92 points lower obesity ($-0.97$ to $-0.87$), 0.73 points lower diabetes ($-0.76$ to $-0.69$), and 0.25 points lower CHD ($-0.26$ to $-0.23$). Smaller inverse associations were observed for AF, stroke/TIA, HF, and PAD. All remained significant after Benjamini--Hochberg correction. HC3 standard errors and exclusion of influential LSOAs produced only small changes.

Demographic adjustment weakened the inverse association for seven outcomes, but made the obesity coefficient  slightly more negative. Subsequent deprivation adjustment generally strengthened the inverse associations. Additional adjustment for NO\textsubscript{x} reduced the walkability coefficient for all eight outcomes, whereas NDVI produced comparatively small changes and slightly strengthened the inverse estimates. Simultaneous adjustment for NO\textsubscript{x} and NDVI produced estimates similar to the NO\textsubscript{x}-adjusted models, with attenuation relative to the primary estimate ranging from approximately 20\% for PAD to 68\% for stroke/TIA (Table~\ref{tab:primary-associations}; Figure~\ref{fig:health-associations}A).

\input{tables/primary_associations}

\subsection{Non-linearity and effect modification}

Natural cubic splines improved fit for seven outcomes (nested-model tests, all $p<0.001$), with no evidence of non-linearity for obesity ($p=0.171$; change in Akaike information criterion, $\Delta\mathrm{AIC}=+0.5$). Improvements varied substantially, with reductions in AIC ranging from 27 for HF to 497 for AF. Quintile models nevertheless showed broadly inverse gradients. Adjusted prevalence generally declined progressively across increasing walkability for CHD, diabetes, HF, hypertension, obesity, and stroke/TIA. AF and PAD reached their lowest adjusted prevalence in the fourth quintile, with only modest attenuation in the highest quintile (Figure~\ref{fig:health-associations}B).

Higher NO\textsubscript{x} weakened the inverse walkability association for all eight conditions after Benjamini--Hochberg correction (obesity $q=0.027$). Higher NDVI strengthened the association for AF, CHD, hypertension, PAD, and stroke/TIA, but weakened it for diabetes and obesity (both $q<0.001$); there was no evidence of modification for HF. Higher deprivation weakened associations for AF, CHD, HF, hypertension, PAD, and stroke/TIA, while diabetes differed, with a stronger inverse gradient at greater deprivation ($q<0.001$).

\input{figures/health_associations}
\subsection{Metropolitan spatial structure and association estimates}

Strong spatial dependence was evident in both the exposure and outcomes. Global Moran's $I$ was 0.95 for walkability and ranged from 0.83 to 0.94 across the eight outcomes (all permutation $p \le 0.001$). Residual Moran's $I$ remained 0.60--0.74 after demographic and deprivation adjustment, indicating substantial residual spatial structure.

Spatial-error modelling substantially reduced the magnitudes of all eight coefficients (Figure~\ref{fig:spatial-dependence}B). For obesity, no clear association remained (relative difference $+0.77$\% per SD; 95\% CI $-0.19$\% to $+1.73$\%). For the other seven outcomes, reductions ranged from 82.8\% for AF to 98.9\% for diabetes. Substantial attenuation also occurred after additional adjustment for NO\textsubscript{x}, NDVI, or both, although its magnitude varied by specification.

Borough fixed effects reduced coefficients by 2--43\%, although all remained inverse and confidence intervals excluded zero (Figure~\ref{fig:spatial-dependence}A). Residual spatial autocorrelation nevertheless remained substantial ($I=0.55$--$0.68$, all permutation $p \le 0.001$). Direct adjustment for centre--periphery structure produced greater attenuation for several outcomes, reducing coefficients by 11.4--75.4\%, with the largest reductions for obesity and CHD and the smallest for PAD. Associations nevertheless remained inverse for all eight outcomes (Table~\ref{tab:spatial-dependence}).

Surrogate surfaces that preserved London's centre--periphery pattern and residual spatial dependence  consistently produced inverse coefficients, reproducing a substantial part of the observed associations. Observed coefficients nevertheless remained more extreme than their corresponding spatially constrained null distributions for all eight outcomes in the primary analyses (all $q\leq0.019$; Figure~\ref{fig:spatial-dependence}C). This distinction persisted for five outcomes after NO\textsubscript{x} adjustment, all eight after NDVI adjustment, and five after joint adjustment. The preserved spatial structure therefore reproduced much, but not all, of the observed alignment.

\subsection{Methodological robustness}

Walkability rankings were highly stable under alternative index definitions. Leave-one-component-out indices showed Spearman correlations of $\rho=0.96$--$0.99$ with the primary equal-weighted index, while an index based on the first principal component had a correlation of $\rho=0.97$. Health associations remained inverse under every alternative, indicating that neither their direction nor the overall pattern of findings depended on any single component or on equal weighting.

The index was similarly stable across 10-, 15-, and 20-minute walksheds ($\rho=0.97$ for 10 versus 15 minutes, $\rho=0.99$ for 15 versus 20 minutes, and $\rho=0.94$ for 10 versus 20 minutes). All health associations remained inverse, and model fit generally improved modestly as walkshed duration increased. The choice of a 15-minute walkshed therefore did not materially alter the direction or inferential conclusions of the health associations. Results were also insensitive to moderate perturbation of the QOF spatial-allocation shares. Across 100 simulations in which within-practice allocation shares were varied by up to $\pm20\%$, all eight coefficients remained inverse and their 95\% confidence intervals excluded zero in every simulation.

Finally, in five-fold spatial-block cross-validation, adding walkability to models containing demographic composition and deprivation increased mean held-out $R^2$ for all eight outcomes (Table~\ref{tab:spatial-dependence}; Figure~\ref{fig:spatial-dependence}D), with increments ranging from 0.025 for PAD to 0.213 for hypertension. The increment was positive in every geographic fold for seven outcomes; PAD showed a negative increment in one fold. However, these predictive analyses do not distinguish walkability from other spatially aligned neighbourhood characteristics.

\input{tables/spatial_dependence}

\input{figures/spatial_dependence}

%% file: tables/index_correlations.tex
\begin{table*}[htbp]
\centering
\caption{Correlations between the composite walkability index, its five components, and two external London walkability measures at LSOA level ($n=4{,}988$--$4{,}994$).}
\label{tab:index-correlations}

\begin{threeparttable}
\small

\begin{tabularx}{\textwidth}{@{}ll*{6}{>{\centering\arraybackslash}X}@{}}
\toprule
\textbf{External index}
& \textbf{Coefficient}
& \textbf{Amenity}
& \textbf{Dwelling}
& \textbf{Land use}
& \textbf{Street}
& \textbf{Transport}
& \textbf{Composite} \\
\midrule

\citet{Stockton2016}
& Spearman $\rho$ & 0.64 & 0.70 & 0.26 & 0.54 & 0.64 & 0.71 \\
& Pearson $r$     & 0.64 & 0.67 & 0.26 & 0.55 & 0.63 & 0.71 \\

\addlinespace

Space Syntax 15-min
& Spearman $\rho$ & 0.85 & 0.81 & 0.11 & 0.72 & 0.82 & 0.85 \\
& Pearson $r$     & 0.49 & 0.38 & 0.06 & 0.50 & 0.59 & 0.52 \\

\bottomrule
\end{tabularx}

\begin{tablenotes}[flushleft]
\footnotesize
\item All correlations $p<0.001$. Stockton: Output Area index of \citet{Stockton2016}, area-weighted from 2001 Output Areas to 2021 LSOAs using polygon intersections ($n=4{,}994$; $n=4{,}988$ for street connectivity). Space Syntax: 15-minute walkability index of the Space Syntax UK Integrated Urban Model, street-length-weighted to 2021 LSOAs. The Space Syntax index is strongly right-skewed, which attenuates Pearson relative to Spearman correlations. LSOA: Lower Layer Super Output Area.
\end{tablenotes}
\end{threeparttable}
\end{table*}

%% file: figures/spatial_anatomy.tex
\begin{figure}[t]
    \centering
    \includegraphics[width=\textwidth]{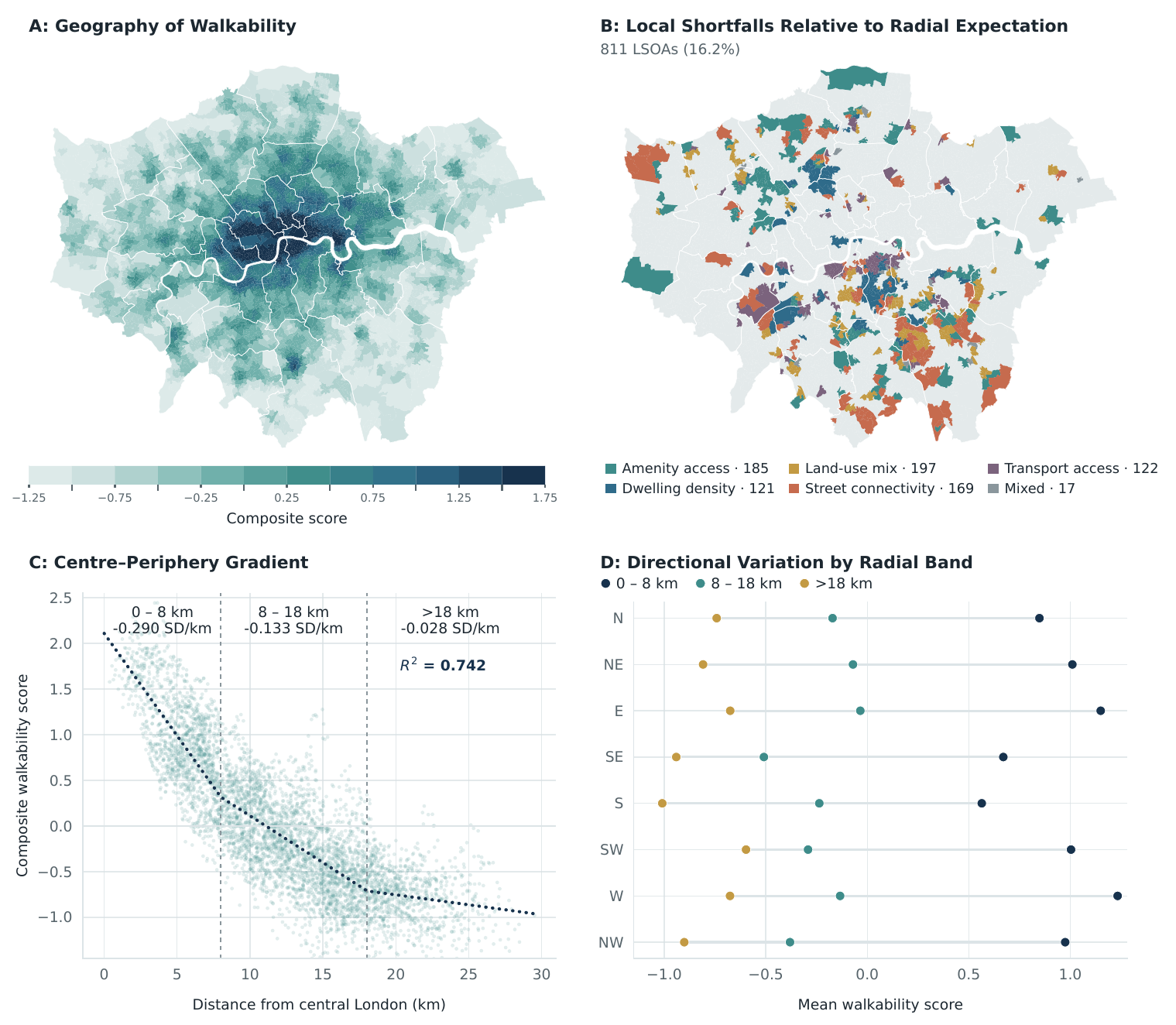}
    \caption{
        Spatial anatomy of walkability across Greater London.
        \textbf{(A)} Geographic distribution of the composite walkability score across LSOAs.
        \textbf{(B)} LSOAs with composite walkability at least one residual standard deviation below that expected from their radial location, classified by the component contributing most strongly to the local shortfall.
        \textbf{(C)} Centre--periphery gradient in composite walkability, showing LSOA-level scores and the fitted piecewise radial trend across 0--8, 8--18, and $>$18~km from central London.
        \textbf{(D)} Directional variation in mean composite walkability across eight compass sectors within the three radial bands.
        LSOA: Lower Layer Super Output Area; SD: standard deviation.
    }
    \label{fig:spatial-anatomy}
\end{figure}

%% file: tables/primary_associations.tex
\begin{table*}[htbp]
\centering
\caption{Associations between neighbourhood walkability and recorded cardiometabolic prevalence under primary and environmental adjustment models.}
\label{tab:primary-associations}

\begin{threeparttable}
\small

\begin{tabularx}{\textwidth}{@{}l*{5}{>{\centering\arraybackslash}X}@{}}
\toprule
Outcome
& Unadjusted
& Primary
& + NO\textsubscript{x}
& + NDVI
& + NO\textsubscript{x} + NDVI \\
\midrule

Hypertension
& -1.82 {\scriptsize(-1.87, -1.78)}
& -1.43 {\scriptsize(-1.48, -1.37)}
& -0.77 {\scriptsize(-0.85, -0.68)}
& -1.51 {\scriptsize(-1.57, -1.45)}
& -0.84 {\scriptsize(-0.93, -0.76)} \\

Obesity
& -0.77 {\scriptsize(-0.82, -0.71)}
& -0.92 {\scriptsize(-0.97, -0.87)}
& -0.55 {\scriptsize(-0.63, -0.47)}
& -0.98 {\scriptsize(-1.03, -0.92)}
& -0.60 {\scriptsize(-0.68, -0.52)} \\

Diabetes
& -0.75 {\scriptsize(-0.79, -0.71)}
& -0.73 {\scriptsize(-0.76, -0.69)}
& -0.40 {\scriptsize(-0.45, -0.34)}
& -0.77 {\scriptsize(-0.80, -0.73)}
& -0.43 {\scriptsize(-0.49, -0.38)} \\

CHD
& -0.38 {\scriptsize(-0.39, -0.37)}
& -0.25 {\scriptsize(-0.26, -0.23)}
& -0.12 {\scriptsize(-0.14, -0.10)}
& -0.25 {\scriptsize(-0.26, -0.24)}
& -0.12 {\scriptsize(-0.14, -0.10)} \\

AF
& -0.29 {\scriptsize(-0.30, -0.28)}
& -0.15 {\scriptsize(-0.16, -0.14)}
& -0.05 {\scriptsize(-0.07, -0.04)}
& -0.15 {\scriptsize(-0.16, -0.14)}
& -0.06 {\scriptsize(-0.08, -0.04)} \\

Stroke/TIA
& -0.18 {\scriptsize(-0.19, -0.18)}
& -0.11 {\scriptsize(-0.12, -0.10)}
& -0.03 {\scriptsize(-0.04, -0.02)}
& -0.11 {\scriptsize(-0.12, -0.11)}
& -0.04 {\scriptsize(-0.05, -0.02)} \\

HF
& -0.09 {\scriptsize(-0.09, -0.09)}
& -0.06 {\scriptsize(-0.07, -0.06)}
& -0.04 {\scriptsize(-0.05, -0.03)}
& -0.06 {\scriptsize(-0.07, -0.06)}
& -0.04 {\scriptsize(-0.05, -0.03)} \\

PAD
& -0.02 {\scriptsize(-0.03, -0.02)}
& -0.02 {\scriptsize(-0.03, -0.02)}
& -0.02 {\scriptsize(-0.02, -0.01)}
& -0.02 {\scriptsize(-0.03, -0.02)}
& -0.02 {\scriptsize(-0.02, -0.01)} \\

\bottomrule
\end{tabularx}

\begin{tablenotes}[flushleft]
\footnotesize
\item Point estimates are percentage-point differences in recorded prevalence per 1-SD increase in walkability. Values in parentheses are 95\% confidence intervals. The primary model adjusts for age composition, sex, ethnicity, and Index of Multiple Deprivation. Environmental models additionally adjust for NO\textsubscript{x}, NDVI, or both. All primary-model associations remained significant after Benjamini--Hochberg correction ($q<0.001$). CHD: coronary heart disease; AF: atrial fibrillation; TIA: transient ischaemic attack; HF: heart failure; PAD: peripheral arterial disease; SD: standard deviation; NO\textsubscript{x}: nitrogen oxides; NDVI: Normalised Difference Vegetation Index.
\end{tablenotes}

\end{threeparttable}
\end{table*}

%% file: figures/health_associations.tex
\begin{figure}[t]
    \centering
    \includegraphics[width=\textwidth]{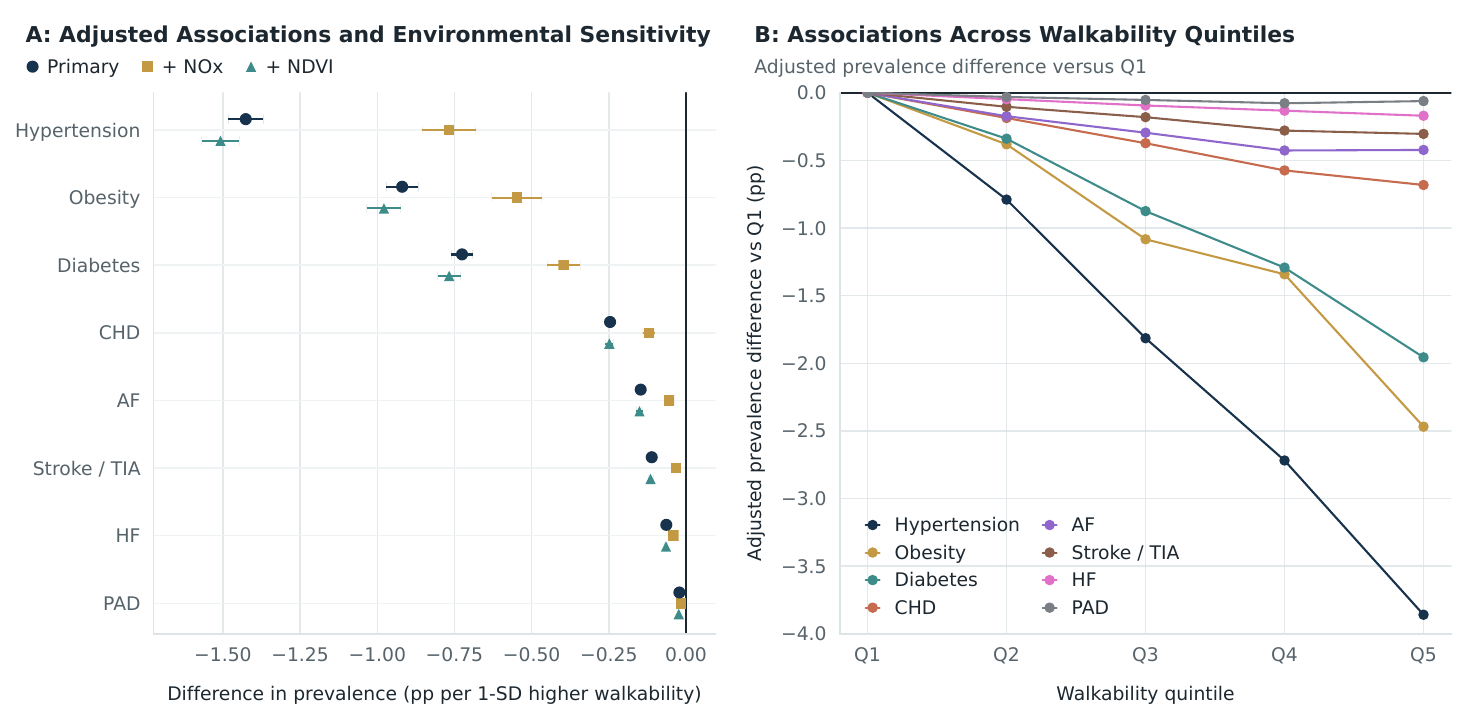}
    \caption{
        Associations between neighbourhood walkability and cardiometabolic morbidity, and their variation across the walkability distribution.
        \textbf{(A)} Estimated differences in recorded prevalence per 1-SD increase in walkability from the primary demographic- and deprivation-adjusted models and from models additionally adjusted for NO\textsubscript{x} or NDVI. Points show regression estimates with 95\% confidence intervals.
        \textbf{(B)} Adjusted prevalence differences across walkability quintiles relative to the least-walkable quintile (Q1). Lines connect estimates from Q1 to Q5 for each cardiometabolic outcome, showing the shape of the relationship across the observed walkability distribution.
        CHD: coronary heart disease; AF: atrial fibrillation; TIA: transient ischaemic attack; HF: heart failure; PAD: peripheral arterial disease; pp: percentage points; SD: standard deviation; NO\textsubscript{x}: nitrogen oxides; NDVI: Normalised Difference Vegetation Index.
    }
    \label{fig:health-associations}
\end{figure}

%% file: tables/spatial_dependence.tex
\begin{table*}[htbp]
\centering
\caption{Sensitivity of walkability--morbidity associations to metropolitan spatial structure.}
\label{tab:spatial-dependence}

\begin{threeparttable}
\small

\setlength{\tabcolsep}{3pt}

\begin{tabular*}{\textwidth}{
    @{\extracolsep{\fill}}
    lcccccc
    @{}
}
\toprule

&
\multicolumn{3}{c}{Absolute Prevalence Models}
& Spatial-Error
& Spatial Null
& Spatial CV \\

\cmidrule(lr){2-4}
\cmidrule(lr){5-5}
\cmidrule(lr){6-6}
\cmidrule(l){7-7}

Outcome
& Primary
& Borough FE
& Centre--Periphery
& Relative \%/SD
& $q$
& $\Delta R^2$ \\

\midrule

Hypertension
& $-$1.43 {\scriptsize(-1.48, -1.37)}
& $-$0.92 {\scriptsize(-0.99, -0.85)}
& $-$0.52 {\scriptsize(-0.60, -0.44)}
& $-$1.36 {\scriptsize[89.8\%]}
& 0.0011
& 0.213 \\

Obesity
& $-$0.92 {\scriptsize(-0.97, -0.87)}
& $-$0.52 {\scriptsize(-0.58, -0.46)}
& $-$0.23 {\scriptsize(-0.30, -0.15)}
& $+$0.77 {\scriptsize[100\%]}
& 0.0011
& 0.128 \\

Diabetes
& $-$0.73 {\scriptsize(-0.76, -0.69)}
& $-$0.49 {\scriptsize(-0.53, -0.44)}
& $-$0.24 {\scriptsize(-0.29, -0.19)}
& $-$0.16 {\scriptsize[98.9\%]}
& 0.0011
& 0.112 \\

CHD
& $-$0.25 {\scriptsize(-0.26, -0.23)}
& $-$0.17 {\scriptsize(-0.18, -0.15)}
& $-$0.08 {\scriptsize(-0.10, -0.06)}
& $-$1.81 {\scriptsize[85.6\%]}
& 0.0011
& 0.131 \\

AF
& $-$0.15 {\scriptsize(-0.16, -0.14)}
& $-$0.10 {\scriptsize(-0.11, -0.08)}
& $-$0.06 {\scriptsize(-0.08, -0.05)}
& $-$2.15 {\scriptsize[82.8\%]}
& 0.0011
& 0.046 \\

Stroke/TIA
& $-$0.11 {\scriptsize(-0.12, -0.10)}
& $-$0.07 {\scriptsize(-0.08, -0.06)}
& $-$0.04 {\scriptsize(-0.05, -0.03)}
& $-$1.45 {\scriptsize[85.3\%]}
& 0.0011
& 0.084 \\

HF
& $-$0.06 {\scriptsize(-0.07, -0.06)}
& $-$0.05 {\scriptsize(-0.06, -0.05)}
& $-$0.03 {\scriptsize(-0.04, -0.02)}
& $-$0.63 {\scriptsize[94.2\%]}
& 0.0011
& 0.102 \\

PAD
& $-$0.02 {\scriptsize(-0.03, -0.02)}
& $-$0.02 {\scriptsize(-0.03, -0.02)}
& $-$0.02 {\scriptsize(-0.02, -0.01)}
& $-$0.72 {\scriptsize[90.2\%]}
& 0.0190
& 0.025 \\

\bottomrule
\end{tabular*}

\begin{tablenotes}[flushleft]
\footnotesize
\item Primary, borough fixed-effect, and centre--periphery estimates are percentage-point differences in recorded prevalence per 1-SD increase in walkability. Values in parentheses are 95\% confidence intervals. Spatial-error models use log-transformed prevalence and are expressed as relative percentage differences per 1-SD. Values in square brackets give attenuation of the corresponding log-scale coefficient, capped at 100\% when coefficients cross zero. Spatially constrained null $q$ values are Benjamini--Hochberg-adjusted empirical two-sided values from 999 surrogate walkability surfaces. Spatial CV $\Delta R^2$ is the mean increment in held-out $R^2$ from adding walkability in five-fold spatial-block cross-validation. FE: fixed effects; CV: cross-validation; SD: standard deviation; CHD: coronary heart disease; AF: atrial fibrillation; TIA: transient ischaemic attack; HF: heart failure; PAD: peripheral arterial disease.
\end{tablenotes}

\end{threeparttable}
\end{table*}

%% file: figures/spatial_dependence.tex
\begin{figure}[H]
    \centering
    \includegraphics[width=\textwidth]{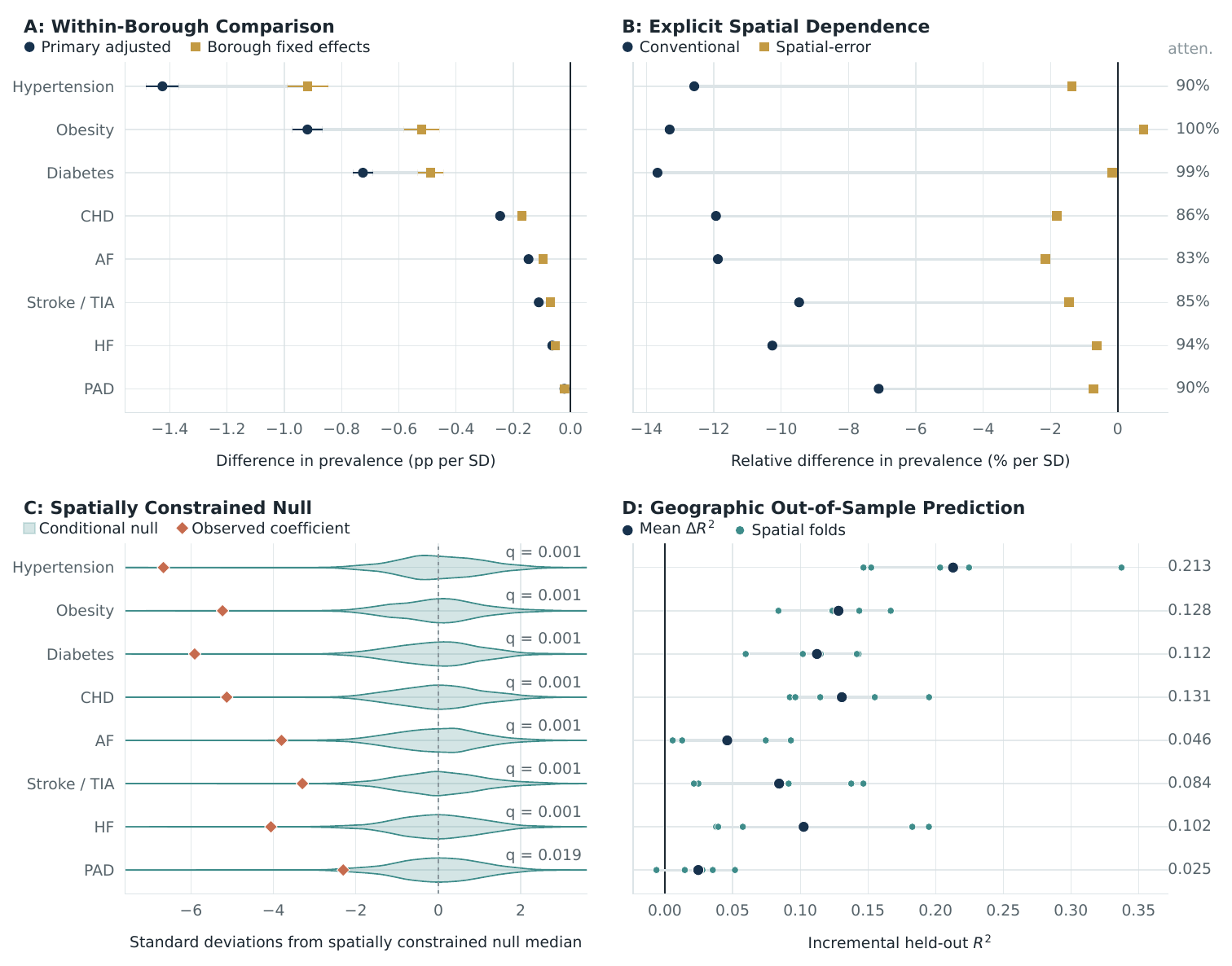}
    \caption{
        Spatial analyses of the association between neighbourhood walkability and cardiometabolic morbidity.
        \textbf{(A)} Comparison of the primary demographic- and deprivation-adjusted estimates with models including borough fixed effects. Points show estimated differences in prevalence per 1-SD increase in walkability with 95\% confidence intervals.
        \textbf{(B)} Conventional and spatial-error estimates from models of log-transformed prevalence, expressed as relative differences in prevalence per 1-SD increase in walkability. Percentages indicate attenuation of the walkability coefficient after modelling residual spatial dependence.
        \textbf{(C)} Observed walkability coefficients compared with spatially constrained null distributions obtained from 999 surrogate walkability surfaces preserving the fitted centre--periphery pattern and residual spatial structure. Estimates are expressed in standard deviations from the corresponding spatially constrained null median.
        \textbf{(D)} Increment in held-out $R^2$ from adding walkability to demographic- and deprivation-adjusted models in five-fold spatial-block cross-validation. Smaller points denote spatial folds and larger points denote the mean across folds.
        CHD: coronary heart disease; AF: atrial fibrillation; TIA: transient ischaemic attack; HF: heart failure; PAD: peripheral arterial disease; pp: percentage points; SD: standard deviation.
    }
    \label{fig:spatial-dependence}
\end{figure}

%% file: sections/discussion.tex
\subsection{Principal findings}

Higher walkability was associated with lower recorded prevalence of all eight cardiometabolic conditions after demographic and socioeconomic adjustment. The largest absolute differences were observed for hypertension, obesity, and diabetes, and associations were generally inverse across the walkability distribution. These findings are consistent with previous studies of walkability and hypertension \citep{Sarkar2018,Jones2021,Chen2024}, obesity and diabetes \citep{Frank2022}, and glucose metabolism \citep{Dales2024}.

Both walkability and morbidity were strongly spatially autocorrelated, with considerable dependence remaining after conventional adjustment. Within-borough comparisons reduced coefficients by 2--43\% and centre--periphery adjustment by 11--75\%. After spatial-error adjustment, no clear association remained for obesity; coefficients for the other seven outcomes were reduced by approximately 83--99\% relative to comparable log-scale models. Attenuation was not uniform across outcomes: it was greatest for obesity under every spatial adjustment and smallest for peripheral arterial disease within boroughs and after centre--periphery adjustment, which may partly reflect differences in how conditions are diagnosed and recorded. Previous work has shown that associations between neighbourhood characteristics and health depend on neighbourhood definition and may vary geographically \citep{Coffee2013,Cebrecos2019,Xu2023}. Our results extend this work by showing that a substantial part of the conventional city-wide association is shared with London's broader metropolitan geography.

The spatially constrained null models did not, however, fully reproduce the observed associations. Surrogate walkability surfaces retaining London's broad centre--periphery pattern, with approximately matched residual dependence, produced substantial inverse associations, but observed coefficients were more extreme for all eight outcomes. This distinction persisted for five outcomes after adjustment for nitrogen oxides. Walkability also improved prediction in areas withheld from model fitting, beyond demographic composition and deprivation alone. These analyses show that the observed local alignment contains information beyond the represented metropolitan structure, but do not establish that walkability, rather than other spatially aligned exposures, explains that alignment.

Walkability itself was predominantly structured by London's centre--periphery pattern. Radial distance explained 74.2\% of variation in the composite index, rising to 80.8\% when directional differences were included. Accessibility, density, connectivity, and public transport formed a common dimension of urban intensity, while land-use mix was comparatively distinct, consistent with its separate contribution to walking behaviour in previous research \citep{Frank2009}. Relative to the radial trend, 16.2\% of neighbourhoods had walkability at least one residual standard deviation below expectation. Absolute rankings therefore obscure some variation between neighbourhoods occupying similar positions within the city. The index also showed correspondence with observed commuting patterns and two existing London walkability measures, while remaining stable across alternative constructions and walkshed durations.

Walkability also coincided with other environmental characteristics relevant to health. Its intensity-related components were positively correlated with nitrogen oxides and negatively correlated with greenness. Adjustment for nitrogen oxides attenuated associations for every outcome, while greenness produced smaller changes. Higher levels of nitrogen oxides generally weakened the inverse walkability association, while the influence of greenness on the association differed between conditions. These findings complement studies examining the joint associations of walkability and air pollution with cardiometabolic health \citep{Howell2019,Zhu2023}. Pollution and greenness are not necessarily independent confounders, however, as they may be correlated exposures or consequences of the same urban development processes. Their inclusion consequently changes the association being estimated rather than simply removing bias.

\subsection{Implications}

The findings have implications for how built-environment associations are studied and interpreted. Previous studies have shown that such associations are sensitive to how neighbourhoods are defined and to the spatial scale of analysis \citep{Coffee2013,Xu2023}, and that walkability measures depend on the spatial unit chosen \citep{Maghelal2011,Shashank2019}. Our findings add the position of a neighbourhood within the metropolitan area as a further source of sensitivity. City-wide comparisons combine variation across London's metropolitan structure, whereas within-borough comparisons focus on variation among neighbourhoods that are closer together. Adjustment for the centre--periphery gradient and spatial-error models address further aspects of broad geographical trends and residual dependence. The differences between their estimates, together with the spatially constrained null models, suggest that shared metropolitan geography should be examined before a city-wide environmental coefficient is interpreted as a local neighbourhood association. Because cities differ in their spatial organisation, this may also be one reason why reported walkability--health associations vary across populations and settings \citep{Chandrabose2019,Kartschmit2020,Cerin2024} and are weaker or absent in some studies \citep{Meijer2025}. The issue is unlikely to be specific to walkability, since other urban characteristics follow similar centre--periphery gradients \citep{Scepanovic2021,Dietz2025Examining}. Spatial analysis should therefore inform study design and interpretation, rather than serving only as a robustness check for conventional estimates.

For urban planning, the findings highlight the importance of distinguishing broad city-wide patterns from local deficiencies in walkability. Neighbourhood design is increasingly considered in relation to cardiometabolic health \citep{Koohsari2020,Adhikari2021,Frank2022}, and walkability indices produced over large areas describe neighbourhoods by their absolute scores \citep{Thomas2017,Herrmann2019,Lam2022}. London's pronounced centre--periphery gradient, documented previously \citep{Stockton2016}, means that neighbourhoods with low absolute scores are not necessarily poorly served relative to their geographical context. Comparing neighbourhoods with the wider spatial pattern can distinguish these broad structural differences from local shortfalls, while examining individual components reveals the characteristics underlying them. This distinction is important because the dimensions of walkability influence walking through distinct mechanisms \citep{Ewing2010,Venerandi2024}, so opportunities for improvement vary between neighbourhoods and across components. Public-transport provision and access to amenities may require investment and coordination across wider areas, whereas deficiencies in street connectivity may offer more localised opportunities for intervention. The close relationships between walkability, traffic-related pollution, and greenness further emphasise the importance of considering pedestrian accessibility alongside other aspects of environmental quality, given evidence that traffic-related pollution may offset some benefits of active travel \citep{Howell2019,Zhang2025,Zhu2023} and that greenness may provide complementary benefits \citep{Scheer2024,Dzhambov2025}. Together, these findings support a more geographically informed approach to assessing neighbourhood walkability and identifying priorities for urban improvement.

\subsection{Limitations and future research}

The cross-sectional, ecological design prevents inference about individual risk or causal effects. Residential self-selection, individual socioeconomic circumstances, physical activity, and exposure outside the home neighbourhood cannot be adequately addressed with the available data. Longitudinal studies incorporating individual characteristics, residential mobility, and daily activity spaces would help address these limitations, although sufficiently detailed data are difficult to obtain \citep{Shields2021,Venerandi2024}.

Recorded prevalence was allocated from general practices to neighbourhoods according to where each practice's patients live. Although moderate perturbations of allocation shares produced little change, they cannot account for systematic differences in diagnosis, healthcare use, or disease distribution within practices. Moreover, prevalence represents accumulated disease, potentially predating the measured environment. Individual-level health records linked to residential histories would reduce uncertainty from spatial allocation and allow temporal relationships to be examined more directly.

Finally, the index captures characteristics of the built environment that support walking, rather than how individuals actually move through and experience it. Uniform walking speeds, intersection delays, and transport weights cannot reflect differences in mobility, route choice, or walking behaviour between individuals. The index also omits subjective aspects of the pedestrian environment, including perceived safety, comfort, and attractiveness, which may influence whether accessible routes are used. The index was validated against commuting patterns from the 2021 Census, which is the most recent census available but was conducted during the COVID-19 pandemic. Future validation incorporating observed walking behaviour and pedestrian perceptions would help address these limitations. Residual spatial confounding also remains possible, since the spatial models and surrogate surfaces cannot capture every aspect of London's geography. More broadly, replication across cities and longitudinal studies would help establish whether changes in walkability precede changes in cardiometabolic health.

%% file: sections/conclusion.tex
Neighbourhood walkability in London showed pronounced geographical structure and was inversely associated with recorded prevalence of all eight cardiometabolic conditions. These associations remained consistent in direction across alternative index constructions and neighbourhood scales, but their estimated magnitude fell substantially when geographical structure was represented more explicitly. Within-borough comparisons, centre--periphery adjustment, and spatial-error models all produced attenuation, indicating that a substantial part of the conventional association is shared with broader patterns across London. Spatially constrained surrogate surfaces preserving the centre--periphery pattern and residual dependence reproduced a substantial part of the inverse association, but not associations as extreme as those observed. The relationship between walkability and cardiometabolic morbidity therefore extends beyond the geographical structure represented in these analyses, without establishing walkability itself as the causal exposure.

These findings place walkability within a wider urban environment. The pronounced centre--periphery gradient in the index, its correlation with air pollution and greenness, and the distinct contributions of its components show that walkability reflects several overlapping dimensions of urban form rather than a single neighbourhood characteristic. Its value therefore lies not only in summarising broad differences in environments that support walking, but also in providing a framework for understanding how neighbourhoods differ from the wider pattern across London. Walkability remains a useful measure of urban form and neighbourhood variation, but associations with cardiometabolic health should be interpreted alongside the social, environmental, and geographical context in which they occur.